# The sulfur plume in the Horsehead nebula: New detections of $S_2H$, $SH^+$, and $CO^+$

A. Fuente,[1] G. Esplugues,[2] P. Rivière-Marichalar,[2] D. Navarro-Almaida,[1] R. Martín-Doménech,[1] G. M.Muñoz-Caro,[1] A. Sánchez-Monge,[3,4] A. Taillard,[1] H. Carrascosa,[1] J. J. Miranzo-Pastor,[1] A. Tasa-Chaveli,[1] P. Fernández-Ruiz,[1] V. V. Guzmán,[5] J.R. Goicoechea,[6] M. Gerin,[7] and J. Pety[8,7]

[1] Centro de Astrobiología (CAB), CSIC-INTA, Ctra. de Ajalvir, km 4, Torrejón de Ardoz, 28850 Madrid, Spain
[2] Observatorio Astronómico Nacional (OAN, IGN), Alfonso XII, 3, 28014 Madrid, Spain
[3] Institut de Ciències de l'Espai (ICE), CSIC, Campus UAB, Carrer de Can Magrans s/n, E-08193, Bellaterra (Barcelona), Spain
[4] Institut d'Estudis Espacials de Catalunya (IEEC), E-08860, Castelldefels (Barcelona), Spain
[5] Instituto de Astrofísica, Pontificia Universidad Católica de Chile, Av. Vicuña Mackenna 4860, 7820436 Macul, Santiago, Chile
[6] Instituto de Física Fundamental (CSIC). Calle Serrano 121-123, 28006, Madrid, Spain
[7] LUX, Observatoire de Paris, Université PSL, Sorbonne Université, CNRS, 75014, Paris, France
[8] Institut de Radioastronomie Millimétrique, 38406, Saint Martin d'Heres, France

ABSTRACT

Sulfur is essential for life, but its abundance and distribution in the interstellar medium remain uncertain, with over 90% of sulfur undetected in cold molecular clouds. Sulfur allotropes ($S_n$) have been proposed as possible reservoirs, but the only detected interstellar molecule with a disulfide bond is $S_2H$ in the Horsehead Nebula, making the estimation of sulfur chains abundances difficult. Here we present total-power ALMA images of $H_2S$, $S_2H$, $SO_2$, $CO^+$, and $SH^+$ towards the Horsehead nebula. These observations, with unprecedented sensitivity (rms $\sim$ 1.5 mK), provide the first detections of $SH^+$ and $CO^+$ in this region, together with the identification of a new $S_2H$ line. The comparison of the spectroscopic images of $H_2S$, $S_2H$, $SO_2$, $CO^+$ and $SH^+$ shows that the $S_2H$ emission originates from a warm gas layer adjacent to the photodissociation front. The emission peak of $S_2H$ is offset from those of reactive ions such as $SH^+$, $CO^+$, and $SO^+$, suggesting that gas-phase reactions involving $SH^+$ and $H_2S$ are not the dominant formation pathway of $S_2H$. Instead, we propose that $S_2H$ is desorbed from irradiated grain surfaces by non-thermal processes. The $SH^+$ detection indicates that sulfur is not significantly depleted at the UV-irradiated edge of the molecular cloud, arguing against a major refractory sulfur reservoir in the interior of molecular clouds.

## 1. INTRODUCTION

Sulfur is the tenth most abundant element in the Universe and is known to play a significant role in biological systems. Sulfur is found in a wide variety of biomolecules, such as amino acids, nucleic acids, sugars, and vitamins. Along with hydrogen, carbon, oxygen, nitrogen, and phosphorus, sulfur is considered to be one of the six essential elements for life (Ranjan et al. 2022).

Corresponding author: A. Fuente
afuente@cab.inta-csic.es

However, there are still a lot of open questions about sulfur chemistry and how the composition of the sulfur budget evolves during star and planet formation. The sulfur depletion problem, characterized by the unexpected scarcity of gaseous sulfur in dense molecular clouds and star-forming regions, is one of the most significant challenges in astronomy and astrochemistry (Ruffle et al. 1999).

Despite the identification of over 300 molecules in the interstellar medium (ISM), only 33 of these molecules contain sulfur atoms[1]. In diffuse clouds, sulfur is predominantly found in its cationic form and simple molecules, where gas-phase sulfur abundance nearly matches the solar value (S/H$\sim$1.5 $\times$ $10^{-5}$) (Neufeld et al. 2015; Psaradaki et al. 2024). However, in starless cores, the combined gas-phase abundances of observed sulfur species account for less than 1% of the solar abundance (Vastel et al. 2018; Fuente et al. 2019, 2023). While it might be assumed that sulfur is locked in the icy mantles on dust grains, only OCS and $SO_2$ have been detected in interstellar ices, with their abundances contributing less than 5% of the total sulfur (Boogert et al. 2022; McClure et al. 2023), leaving over 90% unaccounted for. Several theories have been proposed to explain this "missing" sulfur (Ruffle et al. 1999; Vidal et al. 2017; Laas & Caselli 2019; Shingledecker et al. 2020; Fuente et al. 2023; Vitorino et al. 2024). One possibility is that sulfur exists as neutral atomic sulfur, which is undetectable under the conditions typical of molecular clouds (see, e.g., Vidal et al. 2017). In fact, Fuente et al. (2024) detected the [S I] 25.254 $\mu$m line towards the Orion bar, probing that this atom is an important sulfur reservoir in this kind of high-UV illuminated photodissociation region(PDR). Yet, this situation might not be extrapolated to other regions where the gas and dust are well shielded from UV radiation and the dust temperature is well below the atomic sulfur sublimation temperature.

Sulfur allotropes and polysulfanes ($S_n$, $H_2S_n$) have also been proposed as semi-refractory sulfur reservoirs in cold regions based on laboratory experiments and chemical models (Jiménez-Escobar & Muñoz Caro 2011; Shingledecker et al. 2020; Fuente et al. 2023; Cazaux et al. 2022; Carrascosa et al. 2024). Moreover, these compounds have been detected in comets (Calmonte et al. 2016) and meteorites (Aponte et al. 2023) proving that they are common in solids in our planetary system. The detection of sulfur allotropes and large polysulfanes (n>1) in the solid phase remains difficult even with the high sensitivity and angular resolution provided by the James Webb Space Telescope (JWST). Taillard et al. (2025) showed that the most stable allotrope, $S_8$, would remain below the sensitivity of the JWST even if all sulfur were locked in this compound within the molecular cloud, which making it very difficult to observationally constrain the amount of sulfur trapped in these compounds.

Some attempts have been made to detect related compounds in the gas phase. Esplugues et al. (2013) determined upper limits of N($S_2H$)<7.1 $\times$ $10^{13}$ cm$^{-2}$ and N($H_2S_2$)<1.6 $\times$ $10^{14}$ cm$^{-2}$, which are more than 100 times lower than N($SO_2$) $\sim$ a few $10^{16}$ cm$^{-2}$, in Orion KL. Martín-Doménech et al. (2016) searched for $S_2H$ and $H_2S_2$ in the gas phase towards the well-known hot corino, IRAS 16293−2422, obtaining 3$\sigma$ upper limits to their abundances of < 3.3 $\times$ $10^{-8}$ and 2.4 $\times$ $10^{-8}$, respectively. The low abundances of gaseous $S_2H$ and $H_2S_2$ in these regions were interpreted as a consequence of the rapid destruction of these species once sublimated in such a warm and dense environments.

Fuente et al. (2017) detected $S_2H$ towards the Horsehead nebula, being the first ever detection in the ISM. Due to its proximity ($\sim$ 400 pc; Anthony-Twarog 1982) and favored geometry, the PDR located at the edge of the Horsehead nebula is an excellent template for low-UV illuminated PDRs (Abergel et al. 2003). The Horsehead is illuminated by a moderate radiation field ($G_0$$\sim$60-186 Mathis field) and it has been extensively studied at different wavelengths unveiling a rich molecular chemistry, abundant in sulfur compounds and complex organic molecules (Goicoechea et al. 2006; Gratier et al. 2013; Guzmán et al. 2013, 2014; Rivière-Marichalar et al. 2019). Fuente et al. (2017) estimated that the abundance of $S_2H$ is 2 to 6 times lower than that of $H_2S$ towards the PDR. Interestingly, they also presented experimental measurements of the desorption yields which turned to be $\sim$1.2$\times$$10^{-3}$ and <1$\times$$10^{-5}$ molecules per incident photon for $H_2S$ and $S_2H$, respectively. The upper limit to the $S_2H$ photodesorption yield suggests that if $S_2H$ is formed on grain surfaces, photo-desorption is not a competitive mechanism to release the $S_2H$ molecules to the gas phase once they are formed. Other desorption mechanisms such as chemical desorption or cosmic rays (CR) sputtering can increase the gaseous $S_2H$ abundance in the Horsehead. Alternatively, $S_2H$ can be formed via gas phase reactions involving gaseous $H_2S$ and the abundant ions $S^+$ and $SH^+$.

This paper presents total-power (TP) ALMA observations of $S_2H$, $SO_2$, $^{13}$CS, SO, $CO^+$, and $SH^+$. These observations provide new insights into the sulfur chemistry and the formation of $S_2H$ in this nebula.

## 2. OBSERVATIONS

This paper is based on TP ALMA observations of the 2023.1.00737.S (PI: A. Fuente) and 2019.1.00558.S (PI: V. Guzmán) projects. The observations of project 2023.1.00737.S were carried out in Band 6. Only TP observations are used in this paper because of the low sensitivity obtained in the observations performed with the Alma Compact Array (ACA). In fact, only 10% of the requested ACA observations were completed, preclud-ing the detection of most species. Total-power maps cover an area of 120"×120" centered on R.A. (J2000) = 05:40:54.17; Dec (J2000) = −02:28:00.0, which is referred to as Ionization Front (IF), hereafter. This position is shifted by +3.4" relative to the HCO-peak as defined by Pety et al. (2012). Fourteen high spectral windows, with channel width of $\sim$ 195 kHz ($\sim$ 0.16 km s $^{-1}$), were located to observe lines of several species including the lines of $S_2H$, $H_2S$, $S_3$, OCS, $^{13}CS$, $CO^+$, and $SO_2$, shown in Table A.1, which are analyzed in this paper. Note that the hyperfine lines of $S_2H$ are heavily blended and cannot be observationally resolved. In addition to 2023.1.00737.S data, we used TP spectroscopic images of the $SH^+$ and $SO^+$ lines of project 2019.1.00558.S as extracted from the ALMA archive. A detailed description of this project is given by Hernández-Vera et al. (2023) where the $HCO^+$ 4→3 and CO 3→2 data were published.

[1] https://cdms.astro.uni-koeln.de/classic/molecules

## 3. RESULTS

We have detected the $S_2H$ 14(0,14)→13(0,13) line in a region centered at the position R.A:(J2000) = 05:40:55.293 Dec(J2000) = −02:27:51.89, which is adjacent to the IF (see Fig. 1) . Lower energy lines of this species were previously detected by Fuente et al. (2017) using WHISPER data and is therefore the most likely carrier of the detected emission at 220.60388 GHz. To verify this line assignment, we searched for molecular lines at $\sim$ 220.603 GHz that could drive a possible misidentification using the Cologne Database for Molecular Spectroscopy (CDMS)[2] (Müller et al. 2001) and the Jet Propulsion Laboratory (JPL) molecular spectroscopy catalog [3] (Pickett et al. 1998). We have not found any reasonable candidate ±1 MHz around this frequency. The most intense lines correspond to the complex compounds $C_3H_7CN$ ($\nu$=220603.47 MHz) and $C_2H_5OOCH$ ($\nu$=220603.20 MHz). These molecules are only expected to be abundant in hot cores. The non-detection of methanol in our data precludes the existence of a hot core towards this position. Moreover, these lines require high excitation conditions ($E_u \sim$ 300 K), difficult to achieve in the shielded molecular gas associated with this low-UV PDR (Hernández-Vera et al. 2023).

When observing in total power, the rejection of the image band in the ALMA receivers is ∼10-15 dB (ALMA technical book[4]). We also checked the mirror frequency (∼232200.8 MHz), looking for a possible intense line that might be detectable in the signal band, leading to a false detection. The closest reasonable candidate line to this frequency is the CCS N=18→17 J=17→10 line (but $\sim$ 1 MHz off from the mirror frequency). This line was not detected at any of the positions of the WHISPER survey down to a sensitivity of rms $\sim$ 11 mK (private communication). We can therefore conclude that the detection of the $S_2H$ 14(0,14)→13(0,13) line is robust. In addition to $S_2H$, we included lines of longer sulfur chains, in particular of $S_3$, in our setup. However, we have not detected any of the searched $S_3$ lines and only upper limits to their intensities were derived (see Table A.2).

This paper also shows the first detections of $SH^+$ and $CO^+$ towards the Horsehead nebula. Moreover, these are the first ever detection of these reactive ions in a low illumination PDR. The reactive ion $CO^+$ was previously detected in the reflection nebula NGC 7023, but the UV field in NGC 7023 is > 10 times higher than in the Horsehead nebula (Fuente et al. 2003). In the case of $CO^+$, we have detected the two $CO^+$ N=2→1 J=3/2→1/2 and N=2→1 J=5/2→3/2 lines, excluding a possible misidentification. The most intense one, lying at 236.062 GHz, is blended with the $^{13}CH_3OH$ 5(-2,4)→4(-2,3) and 5(2,3)→4 2,2) lines. The non-detection of the main isotopologue $CH_3OH$ 5(-1,4)→4(-2,3) line allows us to discard any possible contribution of the $^{13}CH_3OH$ line to the observed intensity at 236.062 GHz. Finally, the detection of $SH^+$ N= 0→1 J= 1→0, F=3/2→1/2, already referred to in Hernández-Vera et al. (2023), is presented here for the first time.

The $H_2S$ 2(2,0)→2(1,1), OCS 18→17, and $CH_3OH$ 5(-1,4)→4(-2,3) have not been detected in our observations. These molecules show intense emission in lower excitation lines as shown in Guzmán et al. (2014) and Rivière-Marichalar et al. (2019). The non-detection of the high excitation lines listed in Table A.1 suggests that their emission mainly comes from the bulk of the molecular cloud at a temperature $\sim$ 15 K, instead of from the warm and dense layers of the PDR.

The integrated intensity maps of the $S_2H$ 14(0,14)→13(0,13), $SO_2$ 4(2,2)→3(1,3), $CO^+$ N=2→1, J=5/2→3/2, $SH^+$ N= 0→1, J= 1→0, F=3/2→1/2 and $SO^+$ J=15/2→13/2 lines are shown in Fig. 1. For comparison, we also show the integrated intensity map of the $H_2S$ 1(1,0)→1(0,1) line as observed with the IRAM 30m telescope and previously published by Fuente et al. (2017). Our observations allow us to identify several clumps highlighted with squares in Fig. 1, and with differentiated chemistry. The green square indicates the location of the clump located farthest from the PDR, where the $H_2S$ 1(1,0) →1 (0,1) line shows the most intense emission. Hereafter, we will refer to this area as "Mol-peak". The yellow square indicates the position where the high excitation $SO_2$ line shows an emission peak, and hereafter we will refer to this area as the "Sulfur-plume". The compact $S_2H$ emission is inside this square. The pink square corresponds is centered on the IF position. The ions $CO^+$, $SO^+$, and $SH^+$ are detected towards it. In addition, we detect a new clump in the southern part of the illuminated cloud edge where the $CO^+$ emission reaches its maximum value. This region is indicated with a white square in Fig. 1, and we will refer to it as "IF2", hereafter. Unfortunately, the maps of $SH^+$ and $SO^+$ do not cover the whole area inside this square.

[2] https://cdms.astro.uni-koeln.de

[3] https://soec.jpl.nasa.gov

[4] https://almascience.nrao-edu/proposing/technical-handbook

We have obtained the square-averaged spectra towards Mol-peak, Sulfur-plume, IF, and IF2, to characterize the emission of these regions. These spectra are shown in Fig. 2. While $SO_2$ and $S_2H$ peak toward the so-called Sulfur-plume, the ions $CO^+$, $SO^+$, and $SH^+$ peak towards the IF, suggesting that they come from a different layer of the PDR. As expected, the emission of these ions comes from the external layers of the PDR (Sternberg & Dalgarno 1995; Ginard et al. 2012; Treviño-Morales et al. 2016; Goicoechea et al. 2017). The reactive ion $CO^+$ is destroyed by reactions with $H_2$ to form $HCO^+$ and $HOC^+$ and its abundance is only significant in the HI/$H_2$ transition where a substantial fraction of the hydrogen is still in atomic phase (see, e.g., Fuente et al. 2003).

## 4. MOLECULAR COLUMN DENSITIES

### 4.1. *$S_2H$*

Despite the plethora of laboratory experiments and theoretical work showing that long sulfur chains should be abundant in the ISM, their detection remains elusive. The detection of $S_2H$ towards the Horsehead is the only one reported in the ISM, thus far. Moreover, $S_2H$ remains the only molecule detected with a disulfide bond. This lack of detections challenges the understanding of the chemical routes to form these species.

We have combined the data reported by Fuente et al. (2017) with the ALMA data presented here to perform the rotational diagram of $S_2H$ towards the IF. It should be noticed that the angular resolution of TP ALMA $S_2H$ image (HPBW=29") is similar to that of the observations reported by Fuente et al. (2017) (HPBW∼ 24"-27"), making the comparison reliable. We obtain that the observations are well explained with $T_{rot}$ = 17±1 K, N($S_2H$) = (1.9 ± 0.3) × $10^{12}$ $cm^{-2}$. This rotation temperature is slightly higher than that estimated previously, ∼13 K, by Fuente et al. (2017). The fit shown in Fig. 4 has difficulties to account for the upper limit to the emission of the b-type transition $S_2H$ 4 (1,4)→5 (0, 5), suggesting possible non-LTE excitation for this molecule. Combining the $H_2S$ 1(1,0)→1(0,1) observations presented by Fuente et al. (2017) with the upper limit to the emission of the $H_2S$ 2( 2,0)→2( 1,1) obtained in this work, and assuming an ortho-to-para ratio of 3, we derive an upper limit to the $H_2S$ rotation temperature towards the IF, $T_{rot}$ < 11 K. This upper limit is consistent with the rotation temperatures around 10 K derived by Rivière-Marichalar et al. (2019) for several sulfur-bearing species. The different spatial distribution of $H_2S$ and $S_2H$ together with the different excitation conditions of these two molecules are consistent with the interpretation that they come from different regions, and the $S_2H$/$H_2S$ abundance ratio derived by Fuente et al. (2017) is not representative of that prevailing in the layer from which the $S_2H$ 14(1,14)→13(0,13) emission is coming. We calculate an upper limit to the $H_2S$ abundance in this high excitation region by using LTE calculations with $T_{rot}$= 20 K (see Table 2). Using this value, we estimate that $S_2H$/$H_2S$ > 10 at the edge of the molecular cloud.

### 4.2. *Other species*

For the rest of the lines in Table A.1, we have calculated the column densities assuming $T_{rot}$=10 K and $T_{rot}$=20 K as lower and upper limits to the real rotation temperature. For that, we have convolved the ALMA images to a common angular resolution of 29.8" and afterwards extracted the average spectra in the rectangles plotted in Fig. 1. The Gaussian fits to the resulting spectra are shown in Table A.2 and the calculated column densities in Table 2. For most cases, this change in rotation temperature results in an uncertainty of a factor of ∼2-3 in the estimated column densities. However, some estimations are very sensitive to this parameter, in particular those of $H_2S$, $S_2H$, OCS and $SO^+$ vary by a factor 5−10 from $T_{rot}$=10 K and $T_{rot}$=20 K. Taking into account the complexity of the region, with gas layers characterized by different excitation conditions along the line of sight (Hernández-Vera et al. 2023), the column densities derived assuming $T_{rot}$=20 K must be understood as an upper limit to the column density in the warm and dense layer from which the emission of $S_2H$ arises. Our results show that there is a group of species, SO, $SO_2$, $S_2H$, and $H_2CO$, which present the highest column densities towards the Sulfur-plume (see also Fig 1 and Fig 2). A second group of molecules including $SH^+$, $SO^+$, and $CO^+$ present higher column densities towards the IF and the IF2 positions. Interestingly, the molecules that present higher column densities towards the Sulfur-plume are thought to be linked to grain surface chemistry. In particular, $SO_2$ is possibly detected in ices (McClure et al. 2023; Rocha et al. 2024). Guzmán et al. (2011) modeled the $H_2CO$ chemistry in the Horsehead and concluded that the formation of $H_2CO$ on the surface of dust grains and subsequent photo-desorption into the gas-phase is needed to account for the $H_2CO$ abundance in the IF. This argues in favor of an abundance enhancement of $S_2H$, SO, and $SO_2$ due to the erosion of icy mantles at this position. On the contrary, the reactive ions $CO^+$, $SH^+$, and $SO^+$ are more abundant towards the IF. These ions are formed in the gas phase in the HI/$H_2$ transition layer of the PDR (Fuente et al. 2003; Treviño-Morales et al. 2016; Zanchet et al. 2019; Goicoechea et al. 2017).

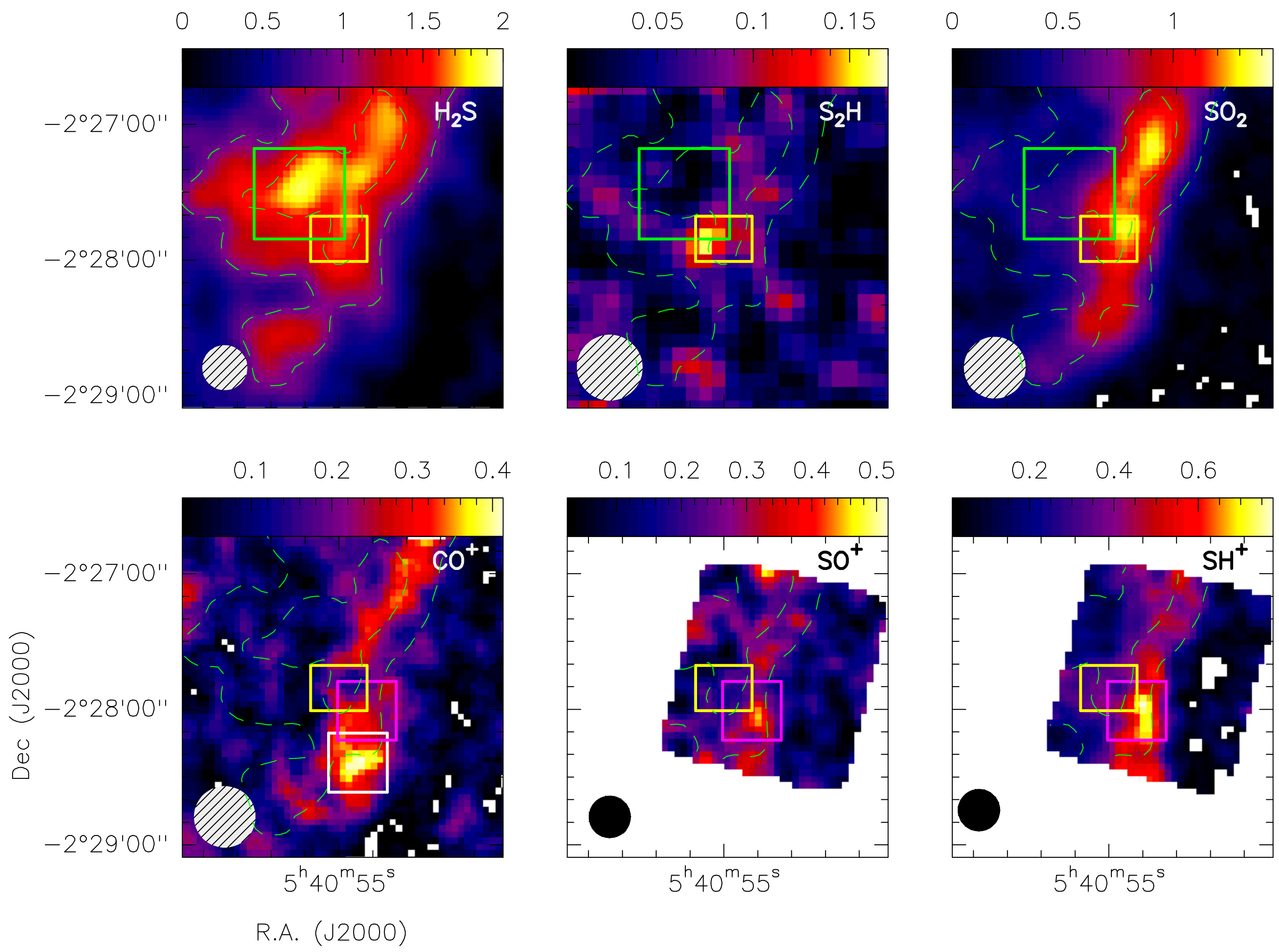

**Figure 1.** Zero-moment images of the $H_2S$ 1(1,0)→1(0,1), $S_2H$ 14(1,14)→13(0,13), $SO_2$ 4(2,2)→3(1,3), $CO^+$ N=2→1, J=5/2→3/2 , $SO^+$ J=15/2→13/2, Ω=1/2, l=e, and $SH^+$ N=0→1, J= 1→0, F=3/2→1/2 lines. The $H_2S$ line is taken from Fuente et al. (2017). Units are K km $s^{-1}$ for $H_2S$ and Jy/beam × MHz for the ALMA maps presented in this work. In colors, we indicate the regions selected to extract the spectra shown in Fig. 2. The beam is drawn in the left-bottom corner. Two contours of the $H_2S$ 1(1,0)→1(0,1) are drawn in all panels for reference.

## 5. DISCUSSION

### 5.1. *Formation of $S_2H$*

Although several evidences point to sulfur allotropes as one of the sulfur reservoirs in the interstellar ices, there is not a reliable chemical network to account for their formation and destruction in astrophysical environments. Shingledecker et al. (2020) predicted that these large compounds can be efficiently formed in interstellar ices, even in the well-shielded regions of molecular clouds, if cosmic-ray driven chemistry and non-diffusive reactions are included. Monte Carlo simulations carried out by Cazaux et al. (2022) showed that the formation of these compounds is also efficient in the translucent region of molecular clouds (Fuente et al. 2019). The destruction mechanisms of these compounds have been less studied, especially for the most stable allotrope,

**Table 2.** Molecular column densities

| Species | Mol. peak | | Sulfur plume | | IF | | IF2 | |
|---|---|---|---|---|---|---|---|---|
| | 10 K ($cm^{-2}$) | 20 K ($cm^{-2}$) | 10 K ($cm^{-2}$) | 20 K ($cm^{-2}$) | 10 K ($cm^{-2}$) | 20 K ($cm^{-2}$) | 10 K ($cm^{-2}$) | 20 K ($cm^{-2}$) |
| (1) | (2) | (3) | (4) | (5) | (6) | (7) | (8) | (9) |
| $H_2S$ ($\times10^{12}$) | <5.90 | <0.26 | <3.08 | <0.14 | <2.71 | <0.12 | <2.62 | <0.12 |
| $S_2H$ ($\times10^{12}$) | <10 | <0.54 | 39.6±6.4 | 2.09±0.33 | 22.0±5.3 | 1.16±0.27 | <27.8 | <1.47 |
| $S_3$ ($\times10^{12}$) | <3.00 | <1.95 | <4.20 | <2.74 | <3.78 | <2.46 | <3.60 | <2.34 |
| OCS ($\times10^{12}$) | <76 | <1.04 | <99 | <1.34 | <91 | <1.24 | <91 | <1.24 |
| SO ($\times10^{12}$) | 10.40±1.0 | 4.50±0.50 | 16.3±1.6 | 7.08±0.71 | 12.9±1.3 | 5.59±0.56 | 9.6±0.96 | 4.15±0.41 |
| $^{13}CS$ ($\times10^{10}$) | 4.22±0.68 | 1.63±0.27 | 4.74±0.80 | 1.83±0.31 | 6.95±1.18 | 2.32±0.40 | <2.00 | <1.00 |
| mirar $SO_2$ ($\times10^{11}$) | 5.40±0.54 | 5.87±0.59 | 9.07±0.91 | 9.85±0.98 | 7.34±0.73 | 8.01±0.80 | 6.40±0.64 | 6.95±0.70 |
| $CO^+$ ($\times10^{10}$) | <0.32* | <0.27* | 1.73±0.19 | 1.42±0.16 | 2.19±0.22 | 1.80±0.18 | 2.18±0.22 | 1.79±0.18 |
| $CH_3OH$ ($\times10^{12}$) | <4.50 | < 0.97 | <5.88 | < 1.27 | <5.78 | < 1.25 | <5.88 | < 1.27 |
| $H_2CO$ ($\times10^{12}$) | 2.50±0.25 | 2.81±0.28 | 2.96±0.29 | 3.32±0.33 | 2.25±0.22 | 2.52±0.25 | 1.78±0.18 | 2.00±0.20 |
| $SH^+$ ($\times10^{11}$) | | | 1.45±0.15 | 1.04±0.10 | 1.66±0.17 | 1.05±0.10 | | |
| $SO^+$ ($\times10^{12}$) | | | 4.18±0.50 | 0.24±0.03 | 4.49±0.45 | 0.25±0.03 | | |

NOTE—* Tentative detection (~3×$\sigma$). The images have been convolved to a common angular resolution of 29.8" to perform column density calculations.

$S_8$. There is also little information on the desorption mechanisms. Several experiments and chemical calculations have provided information on their binding energies (Cazaux et al. 2022; Perrero et al. 2024; Carrascosa et al. 2024) but desorption by UV photons, X-rays, and CR sputtering have been little explored, thus far.

Measurements of the photo-desorption yields of $H_2S$ and $S_2H$ were reported by Fuente et al. (2017). They obtained values of $\sim 1.2 \times 10^{-3}$ and $< 1 \times 10^{-5}$ molecules per incident photon for $H_2S$ and $S_2H$, respectively. The upper limit to the $S_2H$ photodesorption yield suggests that photodesorption is not a competitive mechanism to release the $S_2H$ molecules to the gas phase, opening the possibility that $S_2H$ would be formed in the gas phase. Indeed, $S_2H$ may be produced by the electronic dissociative recombination of $H_2S_2^+$. There are two known $H_2S_2$ gas-phase production pathways from gaseous $H_2S$: the $S^+ + H_2S \rightarrow H_2S_2{}^+ + h\nu$ reaction and the $SH^+ + H_2S \rightarrow H_2S_2{}^+ + H$ reaction. Fuente et al. (2017) proposed that $S_2H$ could be formed in the gas phase after $H_2S$ photo-desorption in the Horsehead nebula.

In this paper, we present the first $S_2H$ image in space. We also present for the first time the image of $CO^+$, $SH^+$ and $SO^+$ in the Horsehead nebula. Our results show that the spatial distribution of $S_2H$ is disjoint with that of $H_2S$ and also with those of $CO^+$, $SH^+$, and $SO^+$. Interestingly, the peak of $S_2H$ is spatially coincident with those of SO, $SO_2$, and $H_2CO$, all of them efficiently produced in hot cores/corinos and bipolar outflows. Although we cannot completely discard the gas phase route, these results suggest that $S_2H$ is formed in UV irradiated ices on the illuminated surface of the molecular cloud, and then released to the gas phase. Based on laboratory experiments, the binding energy of $S_2H$ in $H_2S$ ice is $\sim$ 4264 - 5000 K (Jiménez-Escobar & Muñoz Caro 2011). Assuming that the binding energy in a $H_2O$:$H_2S$ matrix is similar, grain temperatures > 100 K are needed to sublimate this compound. Dust temperatures around 20 - 30 K are expected in this region (Guzmán et al. 2011). Therefore, a non-thermal desorption mechanism such as sputtering by cosmic rays is more likely responsible of the $S_2H$ desorption.

### 5.2. *Sulfur abundance in the Horsehead*

In this section we explore sulfur chemistry using the upgraded version of the Meudon PDR code[5] described in Fuente et al. (2024). This code can simulate very detailed micro-physical processes for a given value of the incident UV field and a 1D density structure, providing as output, the gas and dust temperatures as well as the chemical abundances at each position(Le Petit et al. 2006; Goicoechea & Le Bourlot 2007; Gonzalez Garcia et al. 2008; Le Bourlot et al. 2012; Bron et al. 2014, 2016).

The chemical network includes the basic sulfur and oxygen surface chemistry described in Goicoechea & Cuadrado (2021) to account for sulfur accretion and hydrogenation on grain surfaces. This network is still insufficient to explain the complexity of sulfur chemistry on grain surfaces, ignoring some processes such as the formation of allotropes and complex hydrogen sulfides ($H_2S_n$) that can be efficiently formed in irradiated ices (Cazaux et al. 2022; Carrascosa et al. 2024). Moreover, the model assumes that the chemistry is in steady-state that could not be the case in inner layers ($A_V >$ 2 mag) of the Horsehead nebula (Hernández-Vera et al. 2023). This limits the suitability of the model to explain the $S_2H$ abundance observed towards the Sulfur-plume. This model is, however, the best suited to describe the most external layers of the PDR, $A_V < 2$ mag, where the adsorption of sulfur atoms on grain surfaces is negligible and only gas-phase chemistry is working. The chemistry is expected to be in steady-state in these regions, heavily exposed to the UV field, where chemical timescales are really short. Moreover, the Meudon code includes $H_2$ vibrational state-dependent reaction rates, in particular those computed by Zanchet et al. (2019) for the reaction $S^+ + H_2$ (v) $\rightleftharpoons SH^+ + H$, which is fundamental to account for the formation of $SH^+$ in this low-UV PDR. In these outer layers of the PDR, the ion $SH^+$ is an abundant sulfur species (Goicoechea & Cuadrado 2021) and we can use N($SH^+$)/N($CO^+$) to estimate the sulfur elemental abundance.

[5] https://pdr.obspm.fr

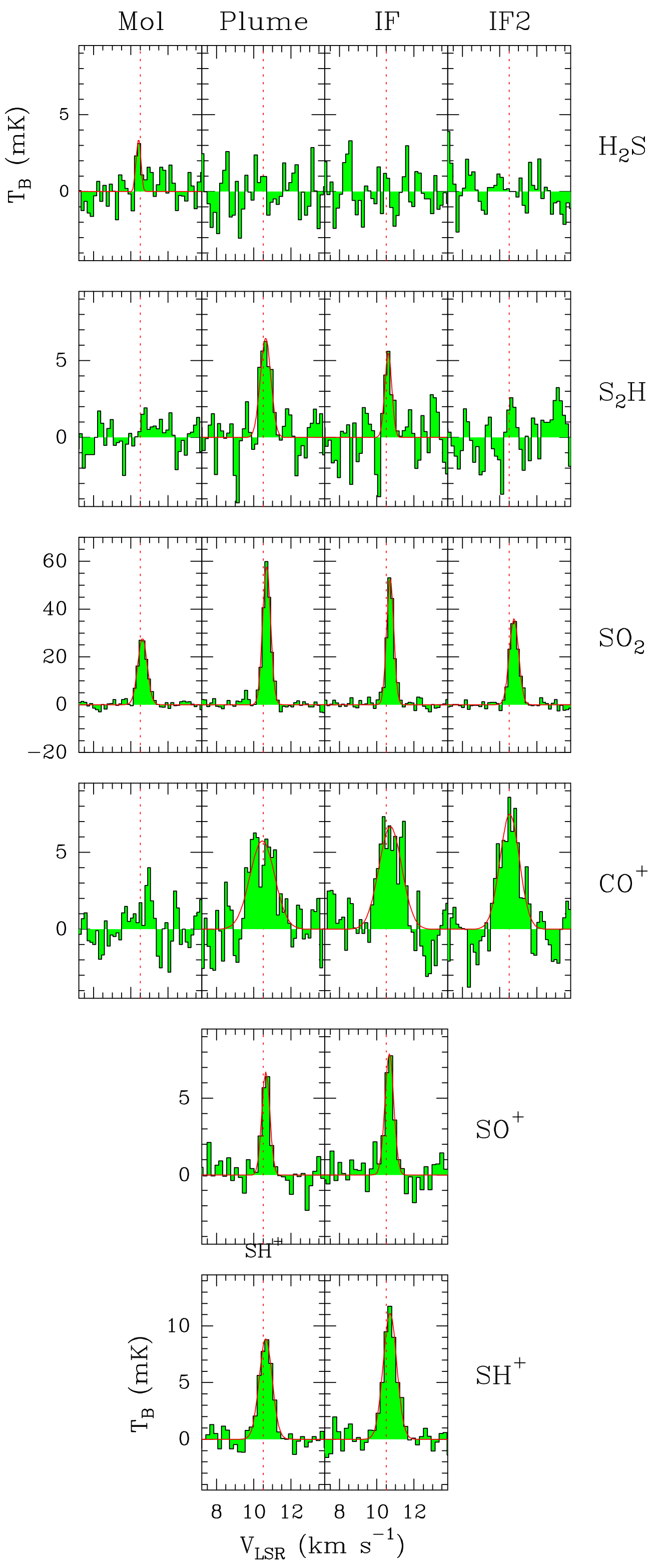

**Figure 2.** Averaged spectra of the $H_2S$ 2(2,0)→2(1,1), $S_2H$ 14(1,14)→13(0,13), $SO_2$ 4(2,2)→3(1,3), $CO^+$ N=2→1, J=5/2→3/2 , $SO^+$ J=15/2→13/2, Ω=1/2, l=e, and $SH^+$ N=0→1, J= 1→0, F=3/2→1/2 lines in the areas defined in Fig. 1.

In Fig. 3, we compare the N($SH^+$)/N($CO^+$) measured towards the IF with the predictions of the Meudon PDR code for 3 different values of sulfur elemental abundance, S/H = 1.5 $\times10^{-5}$, 1.5 $\times10^{-6}$, and 1.5 $\times10^{-7}$. The physical parameters in this simulation are the same as in Hernández-Vera et al. (2023) (see also Table A.3), which successfully reproduce the observations of CO 3→2 and $HCO^+$ 4→3. We have considered two values of $G_0$ = 60, 186 Mathis field in order to check the possible impact of the uncertainties in this parameter on the estimated N($SH^+$)/N($CO^+$) ratio. Although we find variations of the N($SH^+$)/N($CO^+$) ratio at very low visual extinctions, < 1 mag, only moderate values of sulfur depletion, S/H > 1.5 × $10^{-6}$, are compatible with our observations in both cases. These values are in agreement with the results obtained by Goicoechea et al. (2006) and those measured in high-UV PDRs such as the Orion Bar (Goicoechea & Cuadrado 2021; Fuente et al. 2024). This contrasts with the low abundances of sulfur compounds estimated by Rivière-Marichalar et al. (2019). They determined that the sum of the abundances of sulfur species detected towards the IF is ∼ 1.5 $\times10^{-9}$, $\sim10^4$ times lower than the sulfur solar abundance. This suggests the existence of a family of volatile compounds, which remain hidden in millimeter observations, but lock most of the sulfur. Atomic sul-fur is a good candidate to lock an important fraction of sulfur and observations with the JWST can help to constrain its abundance. The detection of $S_2H$ in this nebula, argues in favor of the existence of sulfur chains, which could also lock an important fraction of the sulfur budget.

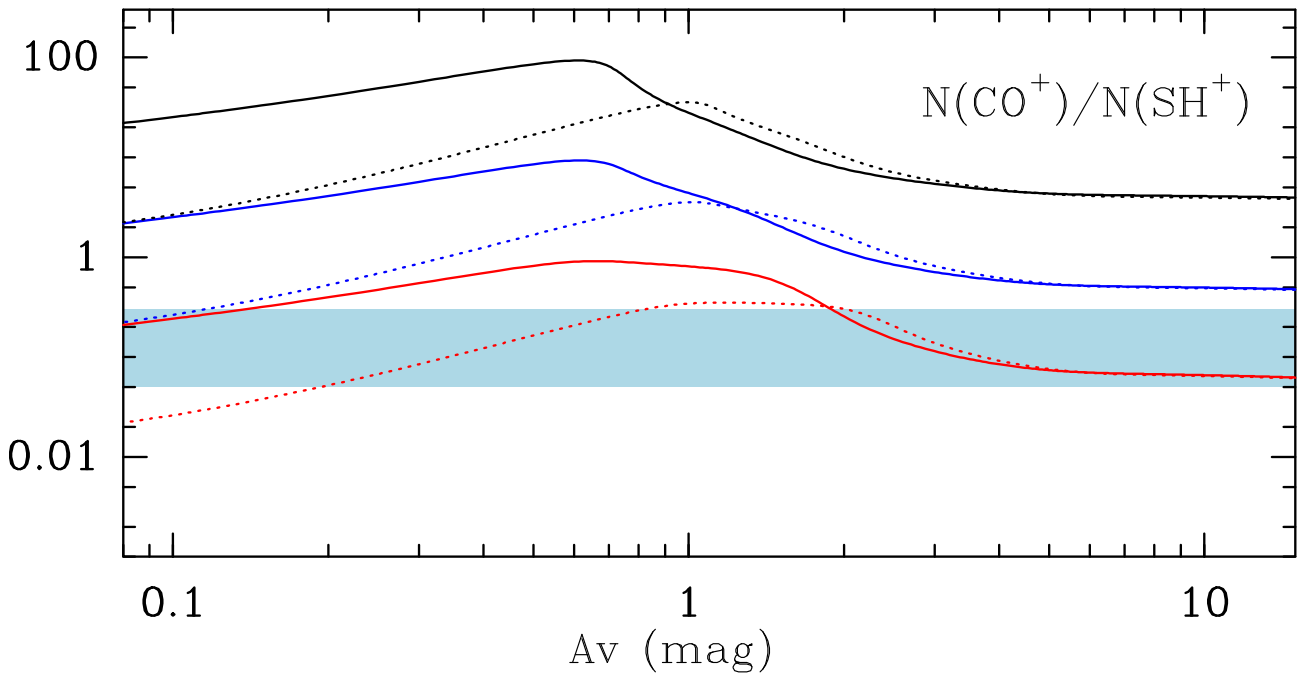

**Figure 3.** The $N(CO^+)/N(SH^+)$ ratio as a function of visual extinction as predicted by the Meudon PDR code using the parameters shown in Table A.3. Different colors correspond to different values of the sulfur elemental abundance: S/H=$1.5\times10^{-5}$ (red), S/H=$1.5\times10^{-6}$ (blue), and S/H=$1.5\times10^{-7}$ (black). Solid and dashed lines correspond to $G_0 = 60$ and 186 Mathis field, respectively. The blue band indicates values of $N(CO^+)/N(SH^+)$=0.08−0.30, consistent with our observations towards the IF assuming an uncertainty of a factor of ∼2. Only S/H > $1.5\times10^{-6}$ can fit our estimations.

## 6. CONCLUSIONS

This paper presents ALMA TP observations of $H_2S$, $S_2H$, $SO_2$, $^{13}CS$, SO, $CO^+$, and $SH^+$, towards the Horsehead nebula, including the first ever shown image of $S_2H$ emission. The detections of $CO^+$ and $SH^+$ are also presented here for the first time. These new detections complement previous observations of $H_2S$ with the IRAM 30m Fuente et al. (2017); Rivière-Marichalar et al. (2019), thus allowing for more detailed study of the sulfur chemistry in the Horsehead nebula. Our results can be summarized as follows:

- The $S_2H$ emission is detected towards the Sulfur-plume and the IF, being maximum towards the former. While $H_2S$ seems to be abundant in the cold molecular gas, as seen from the IRAM-30m observations, the $S_2H$ emission peaks closer to the IF. In the warm layer of the PDR, the $S_2H/H_2S$ abundance ratio could be > 10.

- The reactive ions $CO^+$ and $SH^+$ are detected towards the positions IF and IF2. We have used the Meudon PDR code to estimate the sulfur elemental abundances based on the $N(CO^+)/N(SH^+)$ ratio. Our results are consistent with assuming undepleted sulfur (solar value for the sulfur elemental abundance = $1.5\times10^{-5}$) in the emitting region.

- On the basis of the spatial distribution of targeted species, specially $S_2H$ and $SH^+$, we propose that $S_2H$ is formed on the grain surfaces and released to the gas phase by thermal or non-thermal desorption mechanisms.

We conclude that sulfur elemental abundance is close to the solar value in the PDR associated with the Horsehead nebula. This conclusion is in line with previous works in this same nebula and in the Orion Bar, and points to a more general conclusion about the low depletion or even no depletion at all of sulfur in the regions with intense incident UV flux. Indirectly, this suggests that the sulfur must be in volatiles or semi-refractory material in molecular clouds.

Thus far, a lack of significant sulfur depletion has been observed only in two PDRs, the Orion Bar and the Horsehead nebula. It would be desirable to extend this kind of study to a wide sample of PDRs to further confirm our results and explore possible effects of the environment on the sulfur depletion at the UV-illuminated cloud edges.

This paper makes use of the following ALMA data: ADS/JAO.ALMA#2023.1.00737.S and ADS/JAO.ALMA#2019.1.00558.S. ALMA is a partnership of ESO (representing its member states), NSF (USA) and NINS (Japan), together with NRC (Canada), MOST and ASIAA (Taiwan), and KASI (Republic of Korea), in cooperation with the Republic of Chile. The Joint ALMA Observatory is operated by ESO, AUI/NRAO and NAOJ. This project is co-funded by the European Union (ERC, SUL4LIFE, grant agreement No101096293). AF, PRM, and GE thank project PID2022-137980NB-I00 funded by the Spanish Ministry of Science and Innovation/State Agency of Research MCIN/AEI/10.13039/501100011033 and by "ERDF A way of making Europe". A.S.-M. acknowledges support from the RyC2021-032892-I grant funded by MCIN/AEI/10.13039/501100011033 and by the European Union 'Next GenerationEU'/PRTR, as well as the program Unidad de Excelencia María de Maeztu CEX2020-001058-M, and support from the PID2023-146675NB-I00 (MCI-AEI-FEDER, UE). JRG thanks the Spanish MCINN for funding support under grants PID2023-146667NB-I00. R.M.-D. was supported by a La Caixa Junior Leader grant under agreement LCF/BQ/PI22/11910030.

**Table A.1.** Line list

| Species | Transition | Frequency (GHz) | $\log(A_{ij})$ ($s^{-1}$) | $E_u$ (K) | $g_u$ | HPBW(") |
|---|---|---|---|---|---|---|
| (1) | (2) | (3) | (4) | (5) | (6) | (7) |
| | | 2023.1.00737.S | | | | |
| $H_2S$ | 2(2,0)→2( 1,1) | 216.71044 | -4.31 | 83.98 | 5 | 29.8 |
| $S_3$ | 5(5,1)→4( 4,0) | 218.36217 | -4.89 | 29.42 | 11 | 29.5 |
| OCS | 18→17 | 218.90336 | -4.52 | 99.81 | 37 | 29.5 |
| SO | 6(5)→5(4) | 219.94944 | -3.87 | 34.98 | 13 | 29.3 |
| $S_2H$ | 4(1,4)→5(0,5), J=7/2→9/2, F= 3→4 | 220.49380 | -4.75 | 21.96 | 7 | 29.2 |
| | 4(1,4)→5(0,5), J=7/2→9/2, F= 4→5 | 220.49400 | -4.76 | 21.96 | 9 | 29.2 |
| $S_2H$ | 14(0,14)→13(0,13), J=29/2→27/2, F=14→13 | 220.60385 | -4.09 | 79.39 | 29 | 29.2 |
| | 14(0,14)→13(0,13), J=29/2→27/2, F=15→14 | 220.60388 | -4.09 | 79.39 | 31 | 29.2 |
| $^{13}CS$ | 5→4 | 231.22069 | -3.60 | 33.29 | 22 | 27.9 |
| $S_3$ | 21(6,16)→21(5,17) | 232.38209 | -5.03 | 98.30 | 43 | 27.7 |
| $S_2H$ | 15(1,15)→14(1,14), J=31/2→29/2, F=15→14 | 235.02596 | -4.00 | 103.86 | 31 | 27.4 |
| | 15(1,15)→14(1,14), J=31/2→29/2, F=16→15 | 235.02597 | -4.00 | 103.86 | 33 | 27.4 |
| $SO_2$ | 4(2,2)→3(1,3) | 235.15172 | -4.11 | 19.02 | 11 | 27.4 |
| $CO^+$ | N=2→1, J=3/2→1/2 | 235.789641 | -3.46 | 16.96 | 4 | 27.4 |
| $CO^+$ | N=2→1, J=5/2→3/2 | 236.062553 | -3.38 | 16.99 | 6 | 27.4 |
| $CH_3OH$ | 5(-1,4)→4(-2,3) | 216.94552 | -4.91 | 55.87 | 44 | 29.7 |
| $H_2CO$ | 3(0,3)→2(0,2) | 218.22219 | -3.55 | 20.96 | 7 | 29.5 |
| | | 2019.1.00558.S | | | | |
| $SH^+$ | N= 0→1, J= 1→0, F=3/2→1/2 | 345.94442 | -3.64 | 16.60 | 2 | 18.8 |
| $SO^+$ | J=15/2→13/2, Ω=1/2, l=e | 347.74001 | -3.65 | 70.07 | 16 | 18.7 |

# APPENDIX

## A. SUPPORTING MATERIAL: TABLES AND FIGURES

In the following, we present complementary tables and figures.

**Table A.2.** Gaussian fits to the averaged spectra

| Species | Transition | Area (mK km s$^{-1}$) | V$_{lsr}$ (km s$^{-1}$) | ΔV (km s$^{-1}$) | T (mK) |
|---|---|---|---|---|---|
| (1) | (2) | (3) | (4) | (5) | (6) |
| | | **Mol. peak** | | | |
| $H_2S$ | 2(2,0)→2(1,1) | | rms = 1.21 mK | | |
| $S_3$ | 5(5,1)→4(4,0) | | rms = 1.20 mK | | |
| OCS | 18→17 | | rms = 1.22 mK | | |
| SO | 6(5)→5(4) | 344.83(0.50) | 10.71(0.06) | 0.61(0.01) | 535.01 |
| $S_2H$ | 4(1,4)→5(0,5), J=7/2→9/2 | | rms = 1.10 mK | | |
| $S_2H$ | 14(0,14)→13(0,13), J=29/2→27/2 | | rms = 1.19 mK | | |
| $^{13}CS$ | 5→4 | 4.32(0.69) | 10.61(0.05) | 0.61(0.11) | 6.60 |
| $S_2H$ | 15(1,15)→14(1,14), J=31/2→29/2 | | rms = 1.21 mK | | |
| $SO_2$ | 4(2,2)→3(1,3) | 17.57(0.53) | 10.62(0.01) | 0.60(0.02) | 27.59 |
| $CO^+$ | N=2→1, J=3/2→1/2 | | rms = 1.13 mK | | |
| $CO^+$ | N=2→1, J=5/2→3/2 | 1.78(0.48)* | 10.94(0.06) | 0.43(0.17) | 3.89 |
| $CH_3OH$ | 5(-1,4)→4(-2,3) | | rms = 1.11 mK | | |
| $H_2CO$ | 3(0,3)→2(0,2) | 423.31(0.62) | 0.68(0.06) | 0.68(0.10) | 586.66 |
| | | **Sulfur plume** | | | |
| $H_2S$ | 2(2,0)→2(1,1) | | rms = 1.63 mK | | |
| $S_3$ | 5(5,1)→4(4,0) | | rms = 1.69 mK | | |
| OCS | 18→17 | | rms = 1.55 mK | | |
| SO | 6(5)→5(4) | 542.98(0.71) | 10.74(0.01) | 0.55(0.01) | 920.52 |
| $S_2H$ | 4(1,4)→5(0,5), J=7/2→9/2 | | rms = 1.70 mK | | |
| $S_2H$ | 14(0,14)→13(0,13), J=29/2→27/2 | 4.58 (0.75) | 10.65(0.06) | 0.67(0.12) | 6.37 |
| $^{13}CS$ | 5→4 | 4.85(0.82) | 10.60(0.05) | 0.48(0.10) | 9.47 |
| $S_2H$ | 15(1,15)→14(1,14), J=31/2→29/2 | | rms = 1.70 mK | | |
| $SO_2$ | 4(2,2)→3(1,3) | 29.50(0.64) | 10.69(0.01) | 0.50(0.02) | 54.91 |
| $CO^+$ | N=2→1, J=3/2→1/2 | 5.16(1.10) | 10.82(0.17) | 1.43(0.27) | 3.38 |
| $CO^+$ | N=2→1, J=5/2→3/2 | 9.48(1.04) | 10.48(0.09) | 1.56(0.18) | 5.69 |
| $CH_3OH$ | 5(-1,4)→4(-2,3) | | rms = 1.44 mK | | |
| $H_2CO$ | 3(0,3)→2(0,2) | 500.93(0.73) | 10.67(0.10) | 0.64(0.10) | 739.15 |
| $SH^+$ | N= 0→1, J= 1→0, F=3/2→1/2 | 7.36(0.48) | 10.66(0.03) | 0.87(0.06) | 8.00 |
| $SO^+$ | J=15/2→13/2, Ω=1/2, l=e | 2.96(0.35) | 10.64(0.03) | 0.49(0.06) | 5.60 |
| | | **IF** | | | |
| $H_2S$ | 2(2,0)→2(1,1) | | rms = 1.45 mK | | |
| $S_3$ | 5(5,1)→4(4,0) | | rms = 1.50 mK | | |
| OCS | 18→17 | | rms = 1.45 mK | | |
| SO | 6(5)→5(4) | 429.12(0.66) | 10.77(0.01) | 0.53(0.01) | 756.31 |
| $S_2H$ | 4(1,4)→5(0,5), J=7/2→9/2 | | rms = 1.47 mK | | |
| $S_2H$ | 14(0,14)→13(0,13), J=29/2→27/2 | 2.55(0.59) | 10.62(0.06) | 0.44(0.11) | 5.48 |
| $^{13}CS$ | 5→4 | 6.15(1.05) | 10.71(0.06) | 0.69(0.15) | 8.38 |
| $S_2H$ | 15(1,15)→14(1,14), J=31/2→29/2 | | rms = 1.65 mK | | |
| $SO_2$ | 4(2,2)→3(1,3) | 24.01(0.55) | 10.72(0.01) | 0.45(0.02) | 50.16 |
| $CO^+$ | N=2→1, J=3/2→1/2 | 8.41(1.07) | 10.74(0.10) | 1.35(0.17) | 5.84 |
| $CO^+$ | N=2→1, J=5/2→3/2 | 12.15(1.04) | 10.46(0.06) | 1.38(0.13) | 8.26 |
| $CH_3OH$ | 5(-1,4)→4(-2,3) | | rms = 1.42 mK | | |
| $H_2CO$ | 3(0,3)→2(0,2) | 380.35(0.75) | 10.70(0.01) | 0.65(0.01) | 550.03 |
| $SH^+$ | N= 0→1, J= 1→0, F=3/2→1/2 | 8.42(0.49) | 10.70(0.03) | 0.86(0.06) | 9.24 |
| $SO^+$ | J=15/2→13/2, Ω=1/2, l=e | 3.18(0.32) | 10.65(0.03) | 0.48(0.06) | 6.26 |
| | | **IF2** | | | |
| $H_2S$ | 2(2,0)→2(1,1) | | rms = 1.38 mK | | |
| $S_3$ | 5(5,1)→4(4,0) | | rms = 1.44 mK | | |
| OCS | 18→17 | | rms = 1.44 mK | | |
| SO | 6(5)→5(4) | 318.84(0.72) | 10.80(0.01) | 0.58(0.02) | 513.10 |
| $S_2H$ | 4(1,4)→5(0,5), J=7/2→9/2 | | rms = 1.56 mK | | |
| $S_2H$ | 14(0,14)→13(0,13), J=29/2→27/2 | | rms = 1.52 mK | | |
| $^{13}CS$ | 5→4 | | rms = 1.83 K | | |
| $S_2H$ | 15(1,15)→14(1,14), J=31/2→29/2 | | rms = 1.56 mK | | |
| $SO_2$ | 4(2,2)→3(1,3) | 20.83(0.60) | 10.76(0.01) | 0.57(0.02) | 34.20 |
| $CO^+$ | N=2→1, J=3/2→1/2 | 8.50(0.97) | 10.91(0.07) | 1.23(0.16) | 6.49 |
| $CO^+$ | N=2→1, J=5/2→3/2 | 11.94(0.95) | 10.54(0.06) | 1.43(0.13) | 7.86 |
| $CH_3OH$ | 5(-1,4)→4(-2,3) | | rms = 1.44 mK | | |
| $H_2CO$ | 3(0,3)→2(0,2) | 301.61(0.74) | 10.73(0.01) | 0.69(0.01) | 412.87 |

NOTE—*Doubtful detection. The images have been smoothed to a common angular resolution of 29.8" before extracting the average spectra.

**Table A.3.** Model parameters

| Parameter | Value |
|---|---|
| (1) | (2) |
| $G_0$ | 60, 186 Mathis field |
| $A_V$ | 20 mag |
| Thermal pressure ($P_{th}$) | $3.0 \times 10^6$ K cm$^{-3}$ |
| $R_V$=$A_V$/E(B-V) | 3.1 |
| $N_H$/E(B-V) | $5.80 \times 10^{21}$ cm$^2$ mag$^{-1}$ |
| $\zeta_{H_2}$ | $5 \times 10^{-17}$ s$^{-1}$ |
| C/H | $1.32 \times 10^{-4}$ |
| O/H | $3.19 \times 10^{-4}$ |
| S/H | $1.50 \times 10^{-5}$ |
| N/H | $7.50 \times 10^{-5}$ |

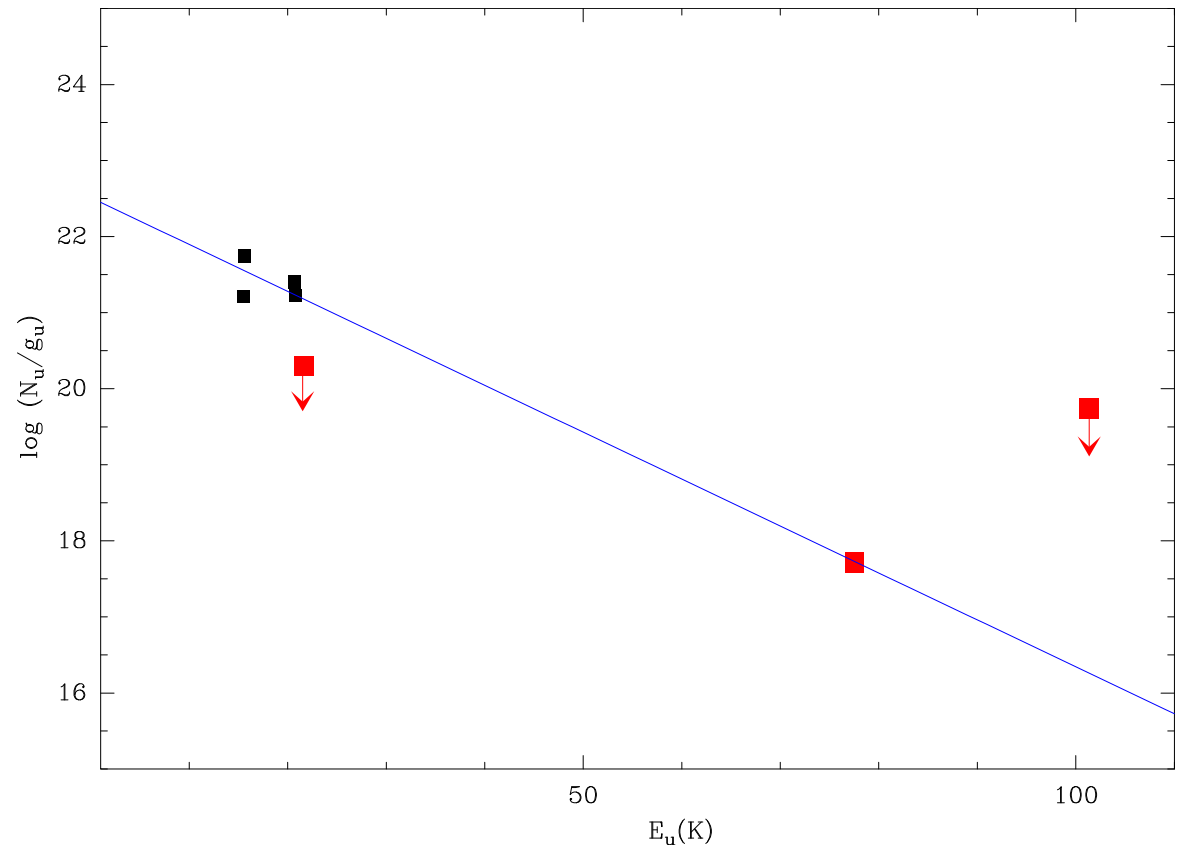

**Figure 4.** Rotational diagram of $S_2H$ based on 30m data from Fuente et al. (2017) (black squares) and the ALMA total power data presented in this work (red squares) assuming a beam-filling factor of 1. It should be noticed that the assumed beam filling factor is not relevant for the calculation of $T_{rot}$ since the observations were performed with similar angular resolution (HPBW ∼ 24"-29").